\documentclass[preprint,amsmath]{aastex}
\newcommand{\etal}{{\it et al.\ }}
\newcommand{\be}{\begin{equation}}
\newcommand{\ee}{\end{equation}}

\newcommand{\ms}{$\mathrm{m \; s}^{-1}$ }

\slugcomment{submitted}
\shorttitle{Planet finding prospects for SIM}
\shortauthors{Ford \& Tremaine}

\begin{document}
\title{Planet finding prospects for the Space Interferometry Mission}
\author{Eric B.\ Ford$^1$, Scott Tremaine$^1$ }

\affil{$^1$Department of Astrophysical Sciences, Princeton University, 
	Peyton Hall, Princeton, NJ 08544-1001, USA}
\email{eford@astro.princeton.edu}

\begin{abstract}
\small
The Space Interferometry Mission (SIM) will make precise astrometric
measurements that can be used to detect planets around nearby stars.
We have simulated SIM observations and estimated the ability of SIM to
detect planets with given masses and orbital periods and measure their
orbital elements.  We combine these findings with an estimate of the
mass and period distribution of planets determined from radial
velocity surveys to predict the number and characteristics of planets
SIM would likely find.  Our predictions are based on extrapolating the
mass distribution of known extrasolar planets by up to a factor of
$\sim100$.  This extrapolation provides the best prediction we can
make of the actual number of planets that SIM will detect and
characterize, but may substantially over- or underestimate the
frequency of Earth-mass planets, especially if these form by a
different mechanism than giant planets. We find that SIM is likely to
detect $\sim1-5$ planets with masses $\le3 M_\oplus$ (depending on
mission parameters).  SIM would measure masses and orbits with 30\%
accuracy for $\sim0-2$ of these planets, but is unlikely to measure
orbits with $10\%$ accuracy for more than one of them.  SIM is likely
to detect $\sim5-25$ planets with mass less than $20 M_\oplus$,
measure masses and orbits with $30\%$ accuracy for $\sim2-12$ of
these, and measure masses and orbits with $10\%$ accuracy for
$\sim2-8$ such planets.  SIM is likely to find $\sim25-160$ planets of
all masses, depending on the observing strategy and mission lifetime.
Roughly $25-65\%$ of the planets detected by SIM have sufficiently
large masses and short orbital periods that they can also be detected
by radial velocity surveys.  Radial velocity surveys could measure
orbital parameters (not including inclination) for $30-70\%$ of the
planets whose orbital parameters will be determined to within $30\%$
by SIM.
\normalsize
\end{abstract}


\section{Introduction}

High precision radial velocity surveys have discovered $\sim100$
planets around nearby main-sequence stars (Butler \etal 2002 and
references therein; \url{http://exoplanets.org}), providing numerous
challenges to theories of planet formation and evolution.  The radial
velocity technique is best suited for massive planets in short-period
orbits around lower main-sequence stars.  Astrometric observations
offer an alternative means for detecting extrasolar planets around
nearby stars.  Since the amplitude of the astrometric perturbation by
a planet increases with the planet's orbital radius and period,
astrometric searches are more sensitive to long-period planets than
radial velocity surveys if their time baseline is long enough.

% \subsection{Motivation}

The Space Interferometry Mission (SIM) is expected to make high
precision targeted observations that will permit the detection of low
mass planets around nearby stars.  Indeed, the detection of low mass
planets is the focus of two of the SIM key projects (Marcy \etal 2002;
Shao \etal 2002) and one of the main science drivers for SIM.  Since
SIM observing time will be limited and very valuable, it is important
to examine possible observing programs and their likely outcomes.  In
this paper, we estimate the number and properties of planets that SIM
is likely to detect for several choices of mission parameters.

%\subsection{SIM Observations}

While SIM is expected to achieve 4 $\mu$as absolute astrometry with
wide-angle observations, it is expected that narrow-angle observations
with $\sim1$ $\mu$as differential astrometry will be most valuable for
planet searches.  In the narrow-angle mode, SIM observations will be
grouped into roughly one hour intervals during which the baseline of
the SIM interferometers will be maintained by locking onto bright
guide stars. During each one hour interval it will be possible to make
up to $\sim10$ one dimensional measurements of the angular separations
of a target star and four reference stars.  Photon noise is smaller
than systematic errors for stars with $V<10.5$.  For pairs of stars
with $V<10.5$, the precision of an individual 30 second observation is
expected to be $\lesssim3$ $\mu$as.  We define the single measurement
precision to be the precision achieved by combining multiple nearly
simultaneous observations in the usual way, assuming that the
observational errors are random and uncorrelated.  Thus, if ten
measurements in a one-hour interval are devoted to a single target
star, then the single-measurement precision is expected to be $\sim3
\mu$as$/\sqrt{10} \sim1 \mu$as.  Given the numerous demands for SIM
observing time, it is expected that the SIM planet searches will
collectively be able to search $\sim$120, 240, or 480 stars with a 1,
1.4, or 2 $\mu$as single measurement precision, respectively
(\url{http://sim.jpl.nasa.gov/ao\_support/index.html}).

Sozzetti \etal (2002) have simulated SIM observations to determine the
probability that SIM would detect a planet given its orbital period
and the amplitude of the astrometric perturbation it induces on its
parent star.  We use their results to determine the ability of SIM to
detect planets around nearby stars, given the actual distances and
masses of stars that SIM is likely to survey.  We randomly assign
planets to these stars based on the observed frequency of extrasolar
planets.  The masses and orbital periods of the simulated planets are
drawn from the planetary mass and period function determined by
Tabachnik \& Tremaine (2002) based on radial velocity observations of
massive planets.

We describe our assumptions and simulations in \S 2.  In \S 3 we
present the results of our simulations.  Based on these results we
answer several questions regarding the discovery potential of SIM in
\S 4.  In \S 5 we summarize our main findings and conclusions.

\section{Methods}

We have conducted Monte Carlo simulations to anticipate the number and
properties of extrasolar planets that the narrow angle SIM planet
search is likely to detect.  First, we generate a list of target stars
appropriate for SIM.  Next, we assign hypothetical planets, specified
by a mass and orbital period or semi-major axis, to these stars using
an empirical fit to the known extrasolar planet mass-period
distribution from Tabachnik \& Tremaine (2002).  Then, we determine
which of these planets SIM would discover using empirical fits to the
detailed simulations of Sozzetti \etal (2002).  Finally, we analyze
the number and the properties of planets discovered to anticipate the
discovery potential of SIM.  For the sake of comparison, we apply a
simplistic model to estimate the number and properties of planets
which could be discovered by a radial velocity survey of the same
stars.

\subsection{Target Stars}

We begin by selecting stars from the Hipparcos catalog within $100$ pc
of the Sun (ESA 1997).  Next, we restrict the sample to stars brighter
than $V = 10.5$.  For dimmer stars the photon noise would dominate the
instrumental uncertainties ($\sim1.7$ $\mu$as) in a 30 second
integration, demanding longer integration times to achieve the
specified astrometric precision. Provided that $V < 10.5$, there is
little gain in astrometric precision or observing efficiency by
observing brighter stars, due to systematic uncertainties
(\url{http://sim.jpl.nasa.gov/ao\_support/index.html}).  We require
that the stars not be a member of a known spectroscopic binary or have
an observed companion within $1''$.  Of the stars which meet the above
criteria, we select $N/4$ stars of each of the four spectral types F,
G, K, and M, for a total of $N$ stars in the sample.  Since the
Hipparcos catalog includes luminosity classifications for only about half of
these stars, we do not exclude giant stars.  We estimate that
$\sim5\%$, and not more than $10\%$ of the stars in our samples are
giant stars.  The actual SIM target list would likely replace these
stars with main-sequence stars at slightly larger distances.

Since obtaining a single measurement precision of $\sigma_d \le 3$
$\mu$as requires averaging the results of multiple 30 second observing
blocks, the necessary observing time scales approximately as $\sim\sigma_d^{-2}$ for
$\sigma_d \le 3$ $\mu$as.  Assuming constant mission time, the number of
stars that can be surveyed therefore scales as $\sim\sigma_d^2$,
provided that there is no additional overhead (e.g., slewing,
acquiring guide stars).  While a target star and its reference stars
must fit within a $\sim1^\circ$ diameter field of view, several target
stars can be observed across a $\sim15^\circ$ diameter field of regard
(FoR) without slewing or significant additional mission time.  Within
a one hour observing interval, the angles between one target star and
four reference stars (all $V<10.5$) may be measured at a precision of
$\sigma_d = 1$ $\mu$as.  Alternatively, in the same hour interval two
or four target stars could be measured at a precision of $\sigma_d =
1.4$ or 2 $\mu$as, respectively, provided that all the target and
reference stars lie within a single FoR
(\url{http://sim.jpl.nasa.gov/ao\_support/index.html}).

\begin{figure}[ht]
\plotone{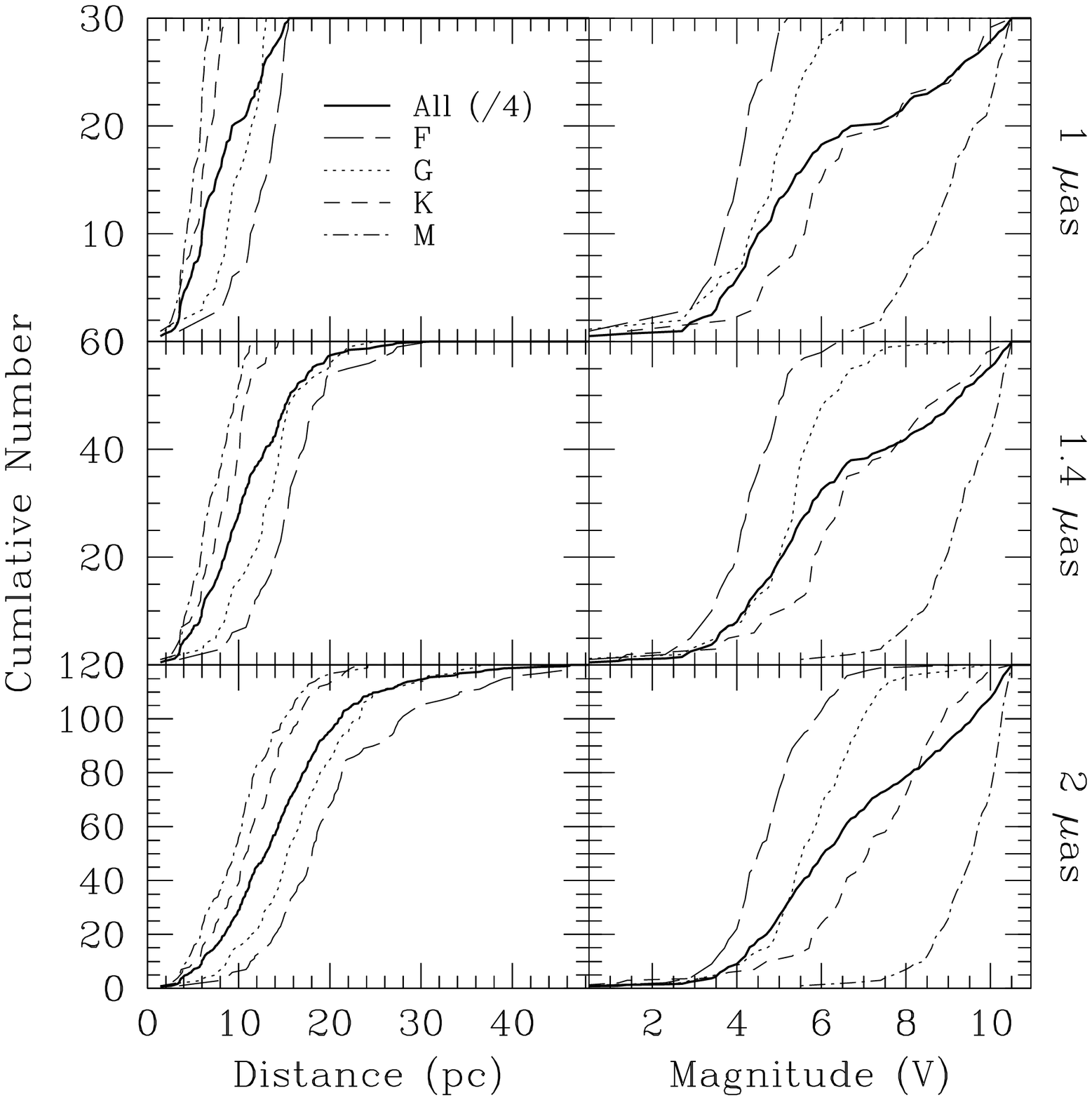}
\caption[Ford.fig1.ps]{
%\small
%
Cumulative number of stars in our target lists as a
function of distance (left) and magnitude (right).
The top, middle, and bottom rows are for the target lists with 120
stars ($\sigma_d =$ 1 $\mu$as), 240 stars ($\sigma_d =$ 1.4 $\mu$as),
and 480 stars ($\sigma_d =$ 2 $\mu$as).
The long dashed, dotted, short dashed, and dashed-dotted lines are
for F, G, K, and M stars, respectively.  The solid bold lines are
for all the stars in each list (the normalization of this line has
been divided by four).
\normalsize
\label{StarProperties}}
\end{figure}

Thus, we take $N=$120, 240, and 480 for $\sigma_d =$1, 1.4, and 2
$\mu$as, respectively.  For the latter two samples, we must also
ensure that one or three additional target stars are within each FoR.
For these target lists we first choose closest candidate stars (the
``primary stars'') and then choose the nearest two or four stars that
satisfy the other criteria above and fit in the required FoRs (the
``secondary stars'').  This requirement accounts for the tails of
distant stars for the 1.4 $\mu$as and 2 $\mu$as target lists shown in
Fig.\ \ref{StarProperties}.  The Hipparcos catalog is incomplete at
magnitudes $\ge 9$, but this incompleteness is only beginning to become significant
for the M stars in the larger target lists.  We will consider a
five-year mission with 24 two dimensional observations per target star and a ten-year
mission with 48 observations per target star.

We also consider an observing strategy similar to that of the
Extrasolar Planet Interferometric Survey (EPIcS) proposed by Shao
\etal (2002). This proposal uses a two-tier approach to observe
$\simeq 75$ of the nearest stars (tier I) at 1 $\mu$as precision 35
times, as well as $\simeq 2100$ nearby stars (tier II) at 4 $\mu$as
precision 25 times (over a five-year mission).  The tier II stars all share
an FoR with one of the tier I stars, making more efficient use of SIM
observation time.  We consider a similar,
but simplified two-tier target list.  To construct a survey that is
directly comparable to the single-tier surveys discussed above, we
choose 52 of the nearest stars (tier I) to observe at 1 $\mu$as
precision and 1092 nearby stars (tier II) at 4 $\mu$as.  Each set of
21 tier II stars must be located within the $15^\circ$ diameter FoR
centered on a tier I star in our simulations.  We observe both the
tier I and tier II stars 24 times during a five-year mission or 48
times during a ten-year mission.  The one-tier and two-tier surveys
which we consider require the same total number of individual 30
second integrations (i.e., $120 \simeq 52 + 1092 / 16$).  While the
EPIcS survey plans to choose tier II targets to ensure a significant
number of several types of stars, for our simulations of a two-tier
survey we do not impose such constraints.  In particular, our
tier II target list includes 49 A stars, 213 F stars, 377 G stars,
354 K stars, and 99 M stars.

% FIGURE 1

\subsection{Planet Model}

Radial velocity surveys have discovered $\sim100$ extrasolar planets
(Butler \etal 2002) and measured their semi-major axis, $a$, and the
ratio of the mass of the planet to the mass of the parent star times
the sine of the orbital inclination of the planet relative to the sky,
$m \sin i / M_\star$, where $M_\star$ is the stellar mass.  The values
of $m \sin i$ for these planets range from $0.12 M_J \simeq 40
M_\oplus$ to over $10 M_J$, where $M_\odot$, $M_J$, and $M_\oplus$ are
the masses of the Sun, Jupiter, and the Earth.  There appears to be a
sharp decrease in the frequency of planets more massive than $\sim10
M_J$.  Tabachnik \& Tremaine (2002) have fit the planet mass
ratio-period function with a power law distribution,
\be
dn = 0.0018 \left(\frac{m}{0.0018 M_\star}\right)^{-0.12}  \left(\frac{P_p}{80 \; \mathrm{d}}\right)^{0.26} \frac{dm}{m} \; \frac{dP_p}{P_p}
\label{EqnTabTre}
\ee
with $dn = 0$ for orbital periods $P_p\le 2$ d or mass ratios $m /
M_\star \ge10^{-2} \simeq 10 M_J / M_\odot$.  We extrapolate this
planet mass function to the smaller masses that will be detectable by
SIM.  We use this functional form to assign planets, their mass
ratios, and their orbital periods to the target stars.  A planet's
semi-major axis is computed from its orbital period using Kepler's
third law, assuming a stellar mass based on the spectral type (Hansen
\& Kawaler 1994).

Note that Tabachnik \& Tremaine (2002) assumed that the stars surveyed for planets all had mass equal to one solar mass.  Thus, an alternative distribution that is equally consistent with their analysis is
\be
dn = 0.0018 \left(\frac{m}{0.0018 M_\odot}\right)^{-0.12}  \left(\frac{P_p}{80 \; \mathrm{d}}\right)^{0.26} \frac{dm}{m} \; \frac{dP_p}{P_p},
\label{EqnTabTreAlt}
\ee
where $M_\star$ in equation (\ref{EqnTabTre}) has been replaced by
$M_\odot$.  If this alternative planet mass function were correct,
then the number of planets of all masses that SIM would detect and
characterize would be slightly increased, but the number of detectable
planets with masses less than $20 M_\oplus$ would be significantly
decreased.

\subsection{SIM Model}

We estimate the ability of SIM to detect a planetary
companion following the simulations of Sozzetti \etal (2002).  They
define $\alpha$, the semi-amplitude of the star's astrometric wobble, by
\be
\frac{\alpha}{''} \equiv \frac{m}{M_\star} \frac{a_p}{\mathrm{AU}} \frac{\mathrm{pc}}{D},
\ee
where $m$ is the mass of the planet, $M_\star$ is the mass of
the parent star, $a_p$ is the semi-major axis of the planet, and
$D$ is the distance to the system from the observer.

Sozzetti \etal (2002) present their results in terms of the
``scaled signal'',
\be
S \equiv \frac{\alpha}{\sigma_d},
\ee
where $\sigma_d$ is the single measurement precision of a set of one
dimensional relative delay measurements between the target star and
each reference star.  Sozzetti \etal (2002) assume $\sigma_d \simeq$ 2
$\mu$as, but their results can be scaled for other values of
$\sigma_d$.  Assuming 24 two-dimensional observations equally spaced
over the mission and an orbital period shorter than the mission
duration, they found that a 50\% probability of detecting a planet
requires $S \ge 1$ and a 95\% probability of detecting a planet
requires $S \ge 2.2$.  The Sozzetti \etal (2002) simulations detect a
planet when the best fit model with no planets can be rejected by a
$\chi^2$ test with 95\% confidence.  Since the target lists that we
consider include 120 to 1144 stars, a 5\% false alarm rate is
unacceptably high.  We prefer to raise the detection threshold of Sozzetti \etal
(2002) by a factor $1.5$, to $S \ge 3.3$, which gives a
negligible false alarm rate ($\simeq0.1\%$), according to our own
simulations.

According to the simulations of Sozzetti \etal (2002), estimating the
mass and orbital parameters to within 30\% of the actual parameters
for 95\% of the planets requires $S \ge 5$.  To measure
the mass of a planet accurate to within 10\% with 95\% probability
requires $S \ge 22$.  To measure the mass and 
orbital parameters to within $10\%$ with $95\%$ probability requires
$S \ge 33$.  

We have conducted our own simulations, which verify the main findings
of Sozzetti \etal (2002).  We agree that $S \ge 33$ is required to
measure the mass and orbital parameters to within $10\%$ with $95\%$
probability.  We also find that the mass and orbital
parameters can be measured to within $10\%$ with $50\%$ probability
when $S \simeq 6$.  Thus, for many planets there is a significant
chance that their orbits will be measured, even for $S$ significantly
smaller than the thresholds given by Sozzetti \etal (2002).  We use
our own simulations to estimate the probability of detecting or
measuring the mass and/or orbital parameters for a given signal $S$
(Ford 2003).  We approximate the probability of detecting a planet or
measuring its mass or orbital parameters based on a fit with the
functional form
\be
P = \left[\frac{1}{1+(S/b)^{-\beta}+(S/c)^{-\gamma}}\right],
\ee
where $b$, $c$, $\beta$, and $\gamma$ are parameters obtained from
fitting to the results of our own simulations and are listed in table
1.  Different values of these fit parameters are used depending on
whether the probability is for detecting a planet, measuring its mass
and orbital parameters with $\le 30\%$ accuracy, measuring its mass
with $\le 10\%$ accuracy, or measuring its mass and orbital parameters
with $\le 10\%$ accuracy.

For orbital periods comparable to or longer than the mission lifetime
($P_{\mathrm{ML}}$), the astrometric signature required for a
detection or measurement increases.  To account for this effect we
have divided our simulations according to the ratio of the planet's
orbital period, $P_p$, to $P_{\mathrm{ML}}$ (see Table 1).  We use
different values of $b$, $c$, $\beta$, and $\gamma$ for each category.
While it is clearly possible to detect the astrometric wobble induced
by a planet with an orbital period much longer than the observational
baseline, our simulations show that it is difficult to measure the
orbital parameters if the observations do not span at least a quarter
of an orbital period.  Thus, we have imposed a sharp upper limit that
prevents measuring orbital parameters --- but allows detections --- of
any planets with orbital periods more than four times the mission
lifetime.  We allow detections for orbital periods up to twelve times
the orbital period; the exact value of this cutoff has little effect on the
results.

We use the above formalism to calculate the probability that a planet
will be detected or have its mass and orbital parameters measured,
given the planet mass ($m$) and orbital period ($P_p$).  Below we will
present results for several sets of mission parameters.  We consider
single measurement precisions of $\sigma_d = 1$, $1.4$, and 2 $\mu$as
for target lists of 120, 240, and 480 stars, respectively.  We
consider a standard mission lifetime $P_{\mathrm{ML}} = 5 \; \mathrm{yr}$
with 24 two dimensional observations of each star as well as an
extended mission lifetime $P_{\mathrm{ML}} = 10 \; \mathrm{yr}$ with 48
two dimensional observations of each star.  We also consider the two-tier
strategy described at the end of \S 2.1.  For the ten-year mission
we reduce the values of $\sigma_d$ by a factor of $\sqrt{2}$.  Our simulations
show that this closely approximates the effect of doubling the number of
observations from 24 to 48.

\subsection{Radial Velocity Model}

To estimate the extent to which planets detected by SIM
will be distinct from those found by radial velocity searches, we
simulate the results of a radial velocity survey for each of our SIM
target lists.  Although precision radial velocity surveys of thousands
of nearby stars are already underway at several ground-based
observatories, often with $50$ or more observations per star, for
simplicity we assume the same number of observations for radial
velocity surveys as for SIM, and that the radial velocity observations
span the same time interval as SIM measurements.  Due to these
assumptions, we expect that we underestimate the number of planets
which would be detected by actual radial velocity surveys.  This is in
contrast to the SIM surveys, for which we have chosen our assumptions
to err on the side of overestimating the number of planets detected.

To estimate which of the planets detected by SIM would likely be
detected by radial velocity surveys, we apply a model similar to that
used for the SIM astrometric survey, but with the scaled signature
being given by
\be
S = \frac{K}{\sqrt{2} \sigma_K},
\label{EqnSRv}
\ee
where $K$ is the star's velocity semi-amplitude, and $\sigma_K \simeq
3\;$ \ms is the accuracy of a single radial velocity measurement.  We
use the same functions to estimate the probability of detecting a
planet or measuring its mass and orbital parameters as we use for SIM.
(Strictly, radial
velocities are sensitive to the combinations $m \sin i$ and
$\omega+\Omega$ rather than the mass, $m$, inclination, $i$, argument
of pericenter, $\omega$, and longitude of ascending node, $\Omega$,
individually.)  We find the use of equation (\ref{EqnSRv})
gives reasonable agreement with the simulations of Cummings \etal
(2002) assuming 24 observations over five years and a 0.1\% false
alarm rate, and are also roughly consistent with the current radial
velocity discoveries (Butler \etal 2002 and references therein).  The
factor of $\sqrt{2}$ in the denominator of the scaled signal for
radial velocity observations accounts for the fact that radial
velocity observations are only sensitive to one component of the
star's velocity, while astrometric observations measure the star's two
dimensional position.  For the ten-year mission we assume 48
observations and reduce the value of $\sigma_K$ by $\sqrt{2}$.

While this analysis neglects many important factors that
affect the efficiency of radial velocity surveys, it allows us to
estimate crudely what fraction of planets found by SIM could be
independently discovered by a radial velocity survey of the same
stars.  Of course, SIM will allow the mass and inclination to be
measured separately which cannot be done with radial velocity
measurements alone.  We do not consider the benefits of combining
radial velocity and astrometric observations (Eisner \& Kulkarni
2002).

\subsection{Summary}

Throughout this paper we attempt to make generous assumptions that
tend to overestimate the number of detections and measurements SIM
will make.  For example, the simulations of Sozzetti \etal (2002)
assume that none of the reference stars have companions and that the
measurement uncertainties are Gaussian and uncorrelated.
Additionally, we assume that multiple planet systems do not reduce
SIM's efficiency for detecting planets and measuring orbits.  Finally,
we allow for detections of planets with orbital periods more than
twice the mission lifetime for which it may be difficult to verify
that the astrometric deviations from a model with no planets are
Keplerian or that they are due to the gravitational perturbation of a
planetary companion.
One change that could significantly increase the number of
planets SIM would find would be that the planetary mass function is
much larger for terrestrial-mass planets than implied by extrapolating
the planetary mass function for giant planets.

\section{Results}
\label{SecResults}

We now present the results of our Monte Carlo simulations.  In
addition to discussing several possible sets of mission parameters, we
frequently divide our results into three categories based on the
planet mass.  Planets with $m < 3 M_\oplus$ would be extremely
interesting since they have a mass comparable to the Earth and are
likely to be rocky bodies formed by a mechanism similar to that which
formed the Earth. This category would include all the terrestrial
planets in our solar system.  We also consider planets with $m < 20
M_\oplus$ that planet formation theories suggest might still be
mainly rocky bodies that have not undergone substantial growth via
gas accretion.  Note that this category would include Uranus and
Neptune as well as the terrestrial planets in our solar system.
Finally, we also present results for planets of all masses, but these
are still limited to $m < 10 M_J$ due to our choice of the planet mass
function.  This category would include all the planets in our solar
system as well as 93\% of the presently known extrasolar planets.

In table \ref{TableStatsAll} we present the number of planets which
are detected or characterized in our simulations for all planets (top),
planets with mass less than $20 M_\oplus$ (middle), and planets with mass less
than $3 M_\oplus$ (bottom).  The different
columns provide the results of simulations with various mission
parameters.  The ranges provided represent 90\% confidence intervals
both in this table and throughout this paper.  These confidence
intervals reflect only statistical uncertainties due to the random
realization of the mass-period distribution (1), not uncertainties
in the distribution itself.

\begin{figure}[ht]
\plotone{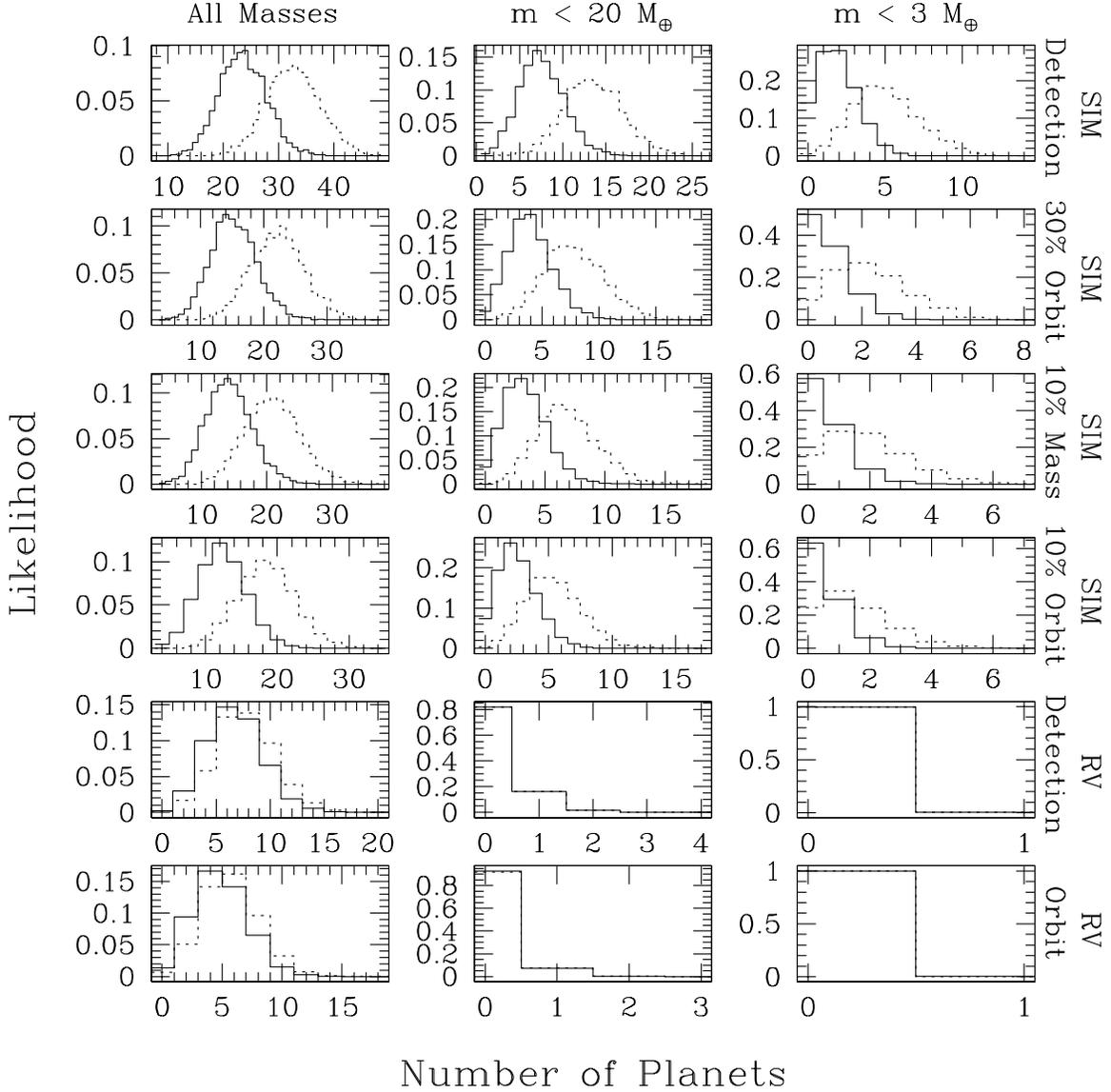}
\caption[Ford.fig2.ps]{
%\small
%
The likelihood of finding a given number of planets.
The top four rows are for a SIM survey with a single measurement
accuracy of 1 $\mu$as that targets 120 F, G, K, and M stars (30 in
each spectral class).  The top row is for detections, the second
row is for estimates of the mass and orbital parameters with 30\% accuracy, the
third row is for mass measurements with 10\% accuracy, and the
fourth row is for measurements of the mass and orbital parameters with 10\%
accuracy.
The bottom two rows are for a radial velocity survey of the
same stars with 3 \ms single measurement accuracy.  The fifth row is
detections and the bottom row is orbital determinations.
The solid lines are for five-year surveys and the dotted lines are for ten-year
surveys.  In some panels the two lines coincide.
The left column is for planets of all masses, the center column is for
planets with mass less than 
$20 M_\oplus$, and the right column is for planets with mass 
less than $3 M_\oplus$.
\normalsize
\label{HistogramGrid}}
\end{figure}

In Fig.\ \ref{HistogramGrid} we show the likelihood of finding a given
number of planets.  The different columns consider planets with
different maximum masses.  The top four rows are for a SIM survey with
1 $\mu$as single measurement precision.  The top row is for detecting
a planet, the second row is for measuring the mass and orbital
parameters with 30\% accuracy, the third row is for measuring the
masses with 10\% accuracy, and the fourth row is for measuring the
masses and orbital parameters with 10\% accuracy.  The bottom two rows are for a
radial velocity survey of the same stars with $3$ \ms single
measurement precision.  The next to bottom row is for detecting the
planet, and the bottom row is for measuring the masses and orbital parameters.
The solid lines are for five-year surveys and the dotted lines are for 10
year surveys.  We conclude that a SIM survey of 120 stars at $1\mu$as
precision is expected to detect $\sim24\pm7$ planets over five years
or $\sim33\pm8$ planets over ten years.  The five-year
survey would measure the masses and orbits of $\sim16\pm6$ of these
planets with 30\% accuracy and $\sim13\pm6$ planets with 10\%
accuracy.  However, most of these planets are relatively massive.  If
we restrict ourselves to planets with mass $\le3M_\oplus$, then even
the ten-year survey would only detect $\sim5\pm3$ planets and would
probably not measure any masses or orbits to 10\% accuracy (see Table
2).

% FIGURE 2

\begin{figure}[ht]
\plotone{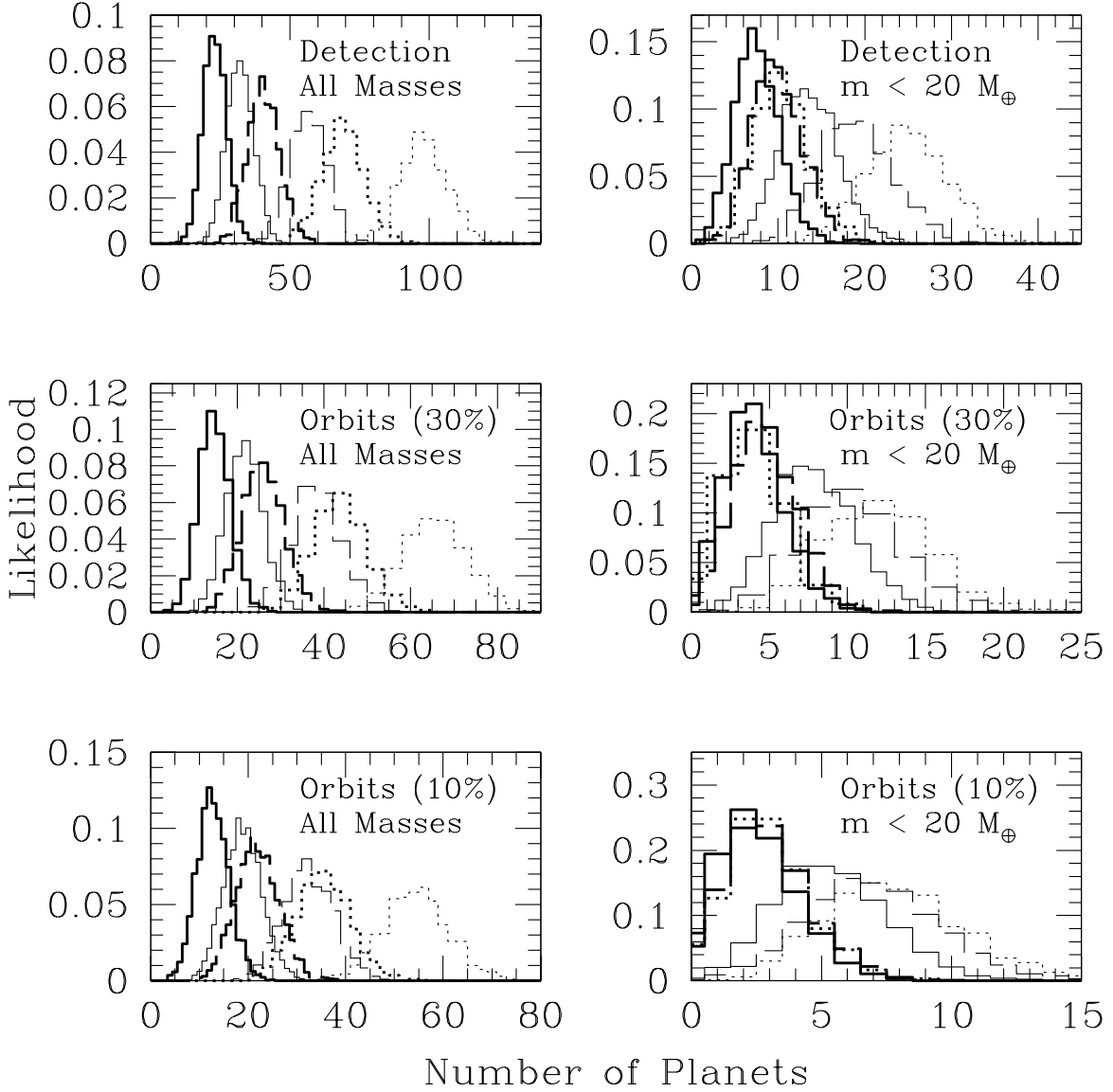}
\caption[Ford.fig3.ps]{
%\small
%
The likelihood of finding a given number of planets.
In each panel, the solid line is for a mission with 1 $\mu$as single
measurement accuracy, the long dashed line for 1.4 $\mu$as, the dotted
line for 2 $\mu$as.  The thick lines are for a five-year mission and
the thin lines are for a ten-year mission.
The top panels are for detecting planets, the middle panels are for
measuring the mass and orbital parameters with 30\% accuracy, and the
bottom panels are for measuring the mass and orbital parameters with
10\% accuracy.  The left panels are for all planets, while the right
panels are for planets with mass less than $20 M_\oplus$.
\normalsize
\label{HistogramMissions}}
\end{figure}

In Fig.\ \ref{HistogramMissions} we explore the effect of varying the
mission parameters.  The solid line is for a mission with 1 $\mu$as
single measurement accuracy, the long dashed line for 1.4 $\mu$as, the
dotted line for 2 $\mu$as.  The thick lines are for a five-year
mission and the thin lines are for a ten-year mission. The upper panel
shows the number of planets detected and the other panels show the
number of planets for which masses and orbits are
determined. The left panels are for all planets, and the right panels
are for planets with $m \le 20 M_\oplus$.  Note that more planets of
all masses are
found by missions targeting a larger number of stars at
lower precision; however, such missions are much less sensitive to the
low-mass terrestrial planets that may be most interesting.

% FIGURE 3

\begin{figure}[pht]
\plotone{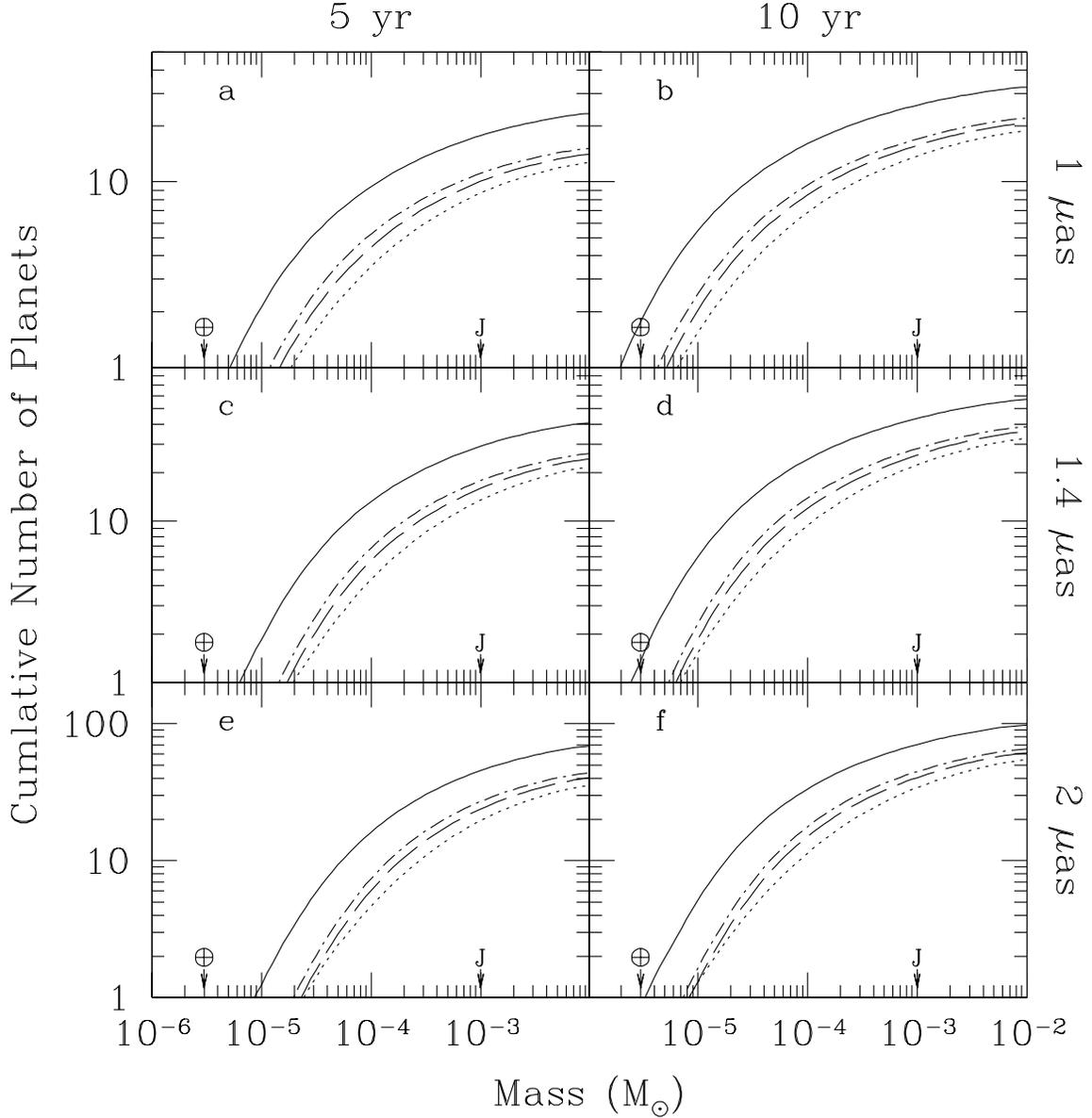}
\caption[Ford.fig4.ps]{
%\small
%
The results of Monte Carlo simulations to estimate
how many planets would be found below a given mass.  We
present the weighted average of thousands of realizations to yield
accurate statistics.
The solid lines are for detections, the dash-dotted lines are for
measurements of the mass and orbital parameters with 30\% accuracy,
the dashed lines are for measurements of the mass with 10\% accuracy, and
the dotted lines are for measurements of the mass and orbital parameters with
10\% accuracy.
Each panel is for a different set of mission parameters and target
lists.  The top panels (a and b) are for 1 $\mu$as single
measurement accuracy and 120 target stars.  The middle panels (c
and d) are for 1.4 $\mu$as single measurement accuracy and 240
target stars.  Bottom panels (e and f) are for 2 $\mu$as single
measurement accuracy and 480 target stars.  The left panels (a, c,
and e) are for a five-year mission, while the right panels (b, d,
and f) are for a ten-year mission.
The masses of Jupiter and the Earth are indicated by J and $\oplus$,
respectively.
\normalsize
\label{CumlativeLevels}}
\end{figure}

\begin{figure}[pht]
\plotone{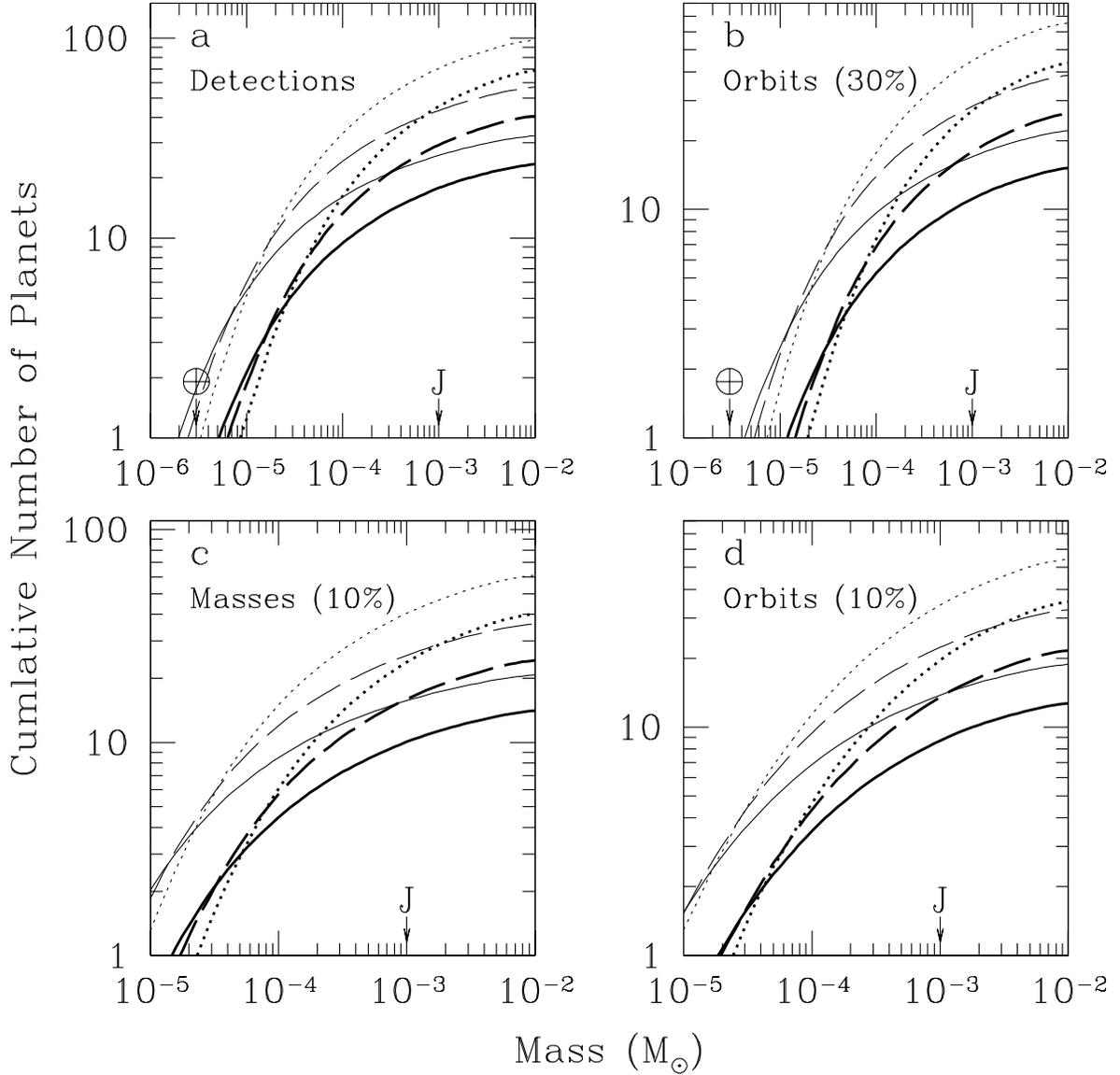}
\caption[Ford.fig5.ps]{
%\small
%
The results of Monte Carlo simulations to estimate
how many planets would be found below a given mass.  We
present the weighted average of thousands of realizations to yield
accurate statistics.
The solid, long dashed, and dotted lines are for missions with a
single measurement accuracy of 1 $\mu$as for 120 target stars, 1.4
$\mu$as for 240 stars, and 2 $\mu$as for 480 stars, respectively.
The thick lines are for a five-year mission and the thin lines for a ten-year
mission.
Panel a is for a detection, panel b for a 30\%  measurement of
the mass and orbital parameters, panel c for a 10\% 
measurement of the mass, and panel d for a 10\% measurement
of the mass and orbital parameters.
\normalsize
\label{CumlativeMissions}}
\end{figure}

In Figs. \ref{CumlativeLevels} and \ref{CumlativeMissions} we plot the
cumulative number of planets found below a given mass. In Fig.
\ref{CumlativeLevels} the different line styles are for detection,
estimating the mass and orbital parameters with 30\% accuracy,
measuring the mass with 10\% accuracy, and measuring the mass and
orbital parameters with 10\% accuracy, while in Fig.\
\ref{CumlativeMissions} the line styles are for surveys with different
single measurement accuracies and hence different numbers of target
stars.  The most effective survey for planets less than a given mass
can be determined by which curve has the largest value for a given
mass in Figure \ref{CumlativeMissions}.  For example, with a five-year
baseline, observing 480 stars with 2 $\mu$as precision would maximize
the total number of planets whose masses are measured with 10\%
accuracy, but all three strategies have a similar yield of planets
with mass $\le 20 M_\oplus$ whose masses would be measured with 10\%
accuracy.

% FIGURE 4

%FIGURE 5 

\begin{figure}[pht]
\plotone{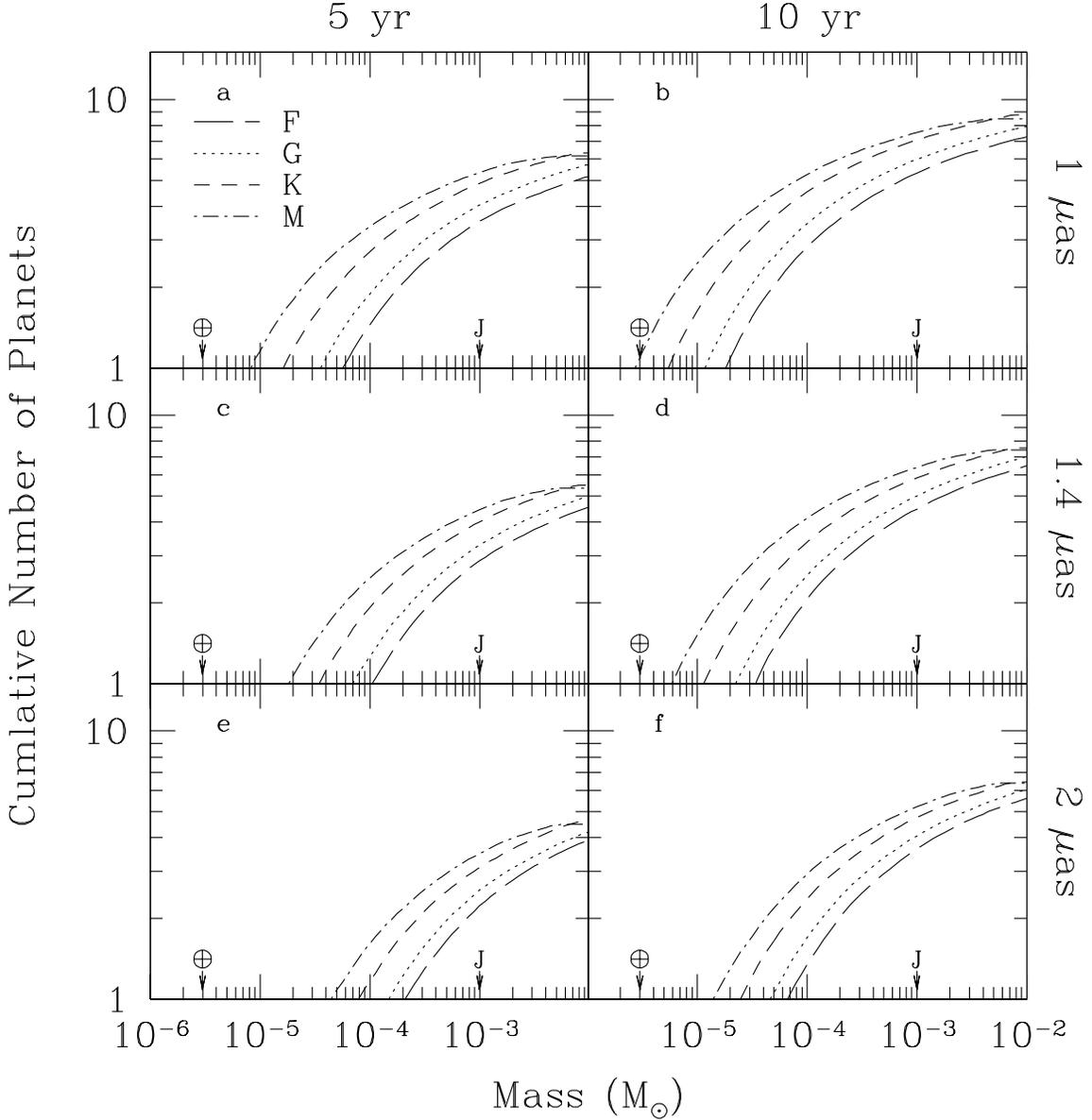}
\caption[Ford.fig6.ps]{
%\small
%
The results of Monte Carlo simulations to estimate
how many planets would be found below a given mass.  We
present the weighted average of thousands of realizations to yield
accurate statistics broken down according to the spectral type of
the star.
The long dashed lines are for F stars, the dotted lines are for G
stars, the short dashed lines are for K stars, and the
dashed-dotted lines are for M stars.
Each panel is for a different set of mission parameters and target
lists.  The top panels (a and b) are for a 1 $\mu$as single
measurement accuracy and 120 target stars equally distributed among
the four spectral types.  The middle panels (c and d) are for a 1.4
$\mu$as single measurement accuracy and 240 target stars.  Bottom
panels (e and f) are for a 2 $\mu$as single measurement accuracy and
480 target stars.
The left panels (a, c, and e) are for a five-year mission, while the
right panels (b, d, and f) are for a ten-year mission.
\normalsize
\label{CumlativeSpectralTypes}}
\end{figure}

In Fig. \ref{CumlativeSpectralTypes} we plot the cumulative number of
planets found below a given mass, divided up by the spectral type of
the star of each planet.  Note that K and M stars are typically closer
and lower mass than F and G stars, increasing the number of planets found around K
and M stars relative to F and G stars.  This effect is more
significant for smaller planetary masses, particularly for masses $\le 10
M_\oplus$.  Note that our assumption that the planet mass function is
given by equation (\ref{EqnTabTre}) rather than equation
(\ref{EqnTabTreAlt}) results in a larger number of detectable low mass
planets around these stars in our simulations.

% FIGURE 6

\begin{figure}[ht]
\plotone{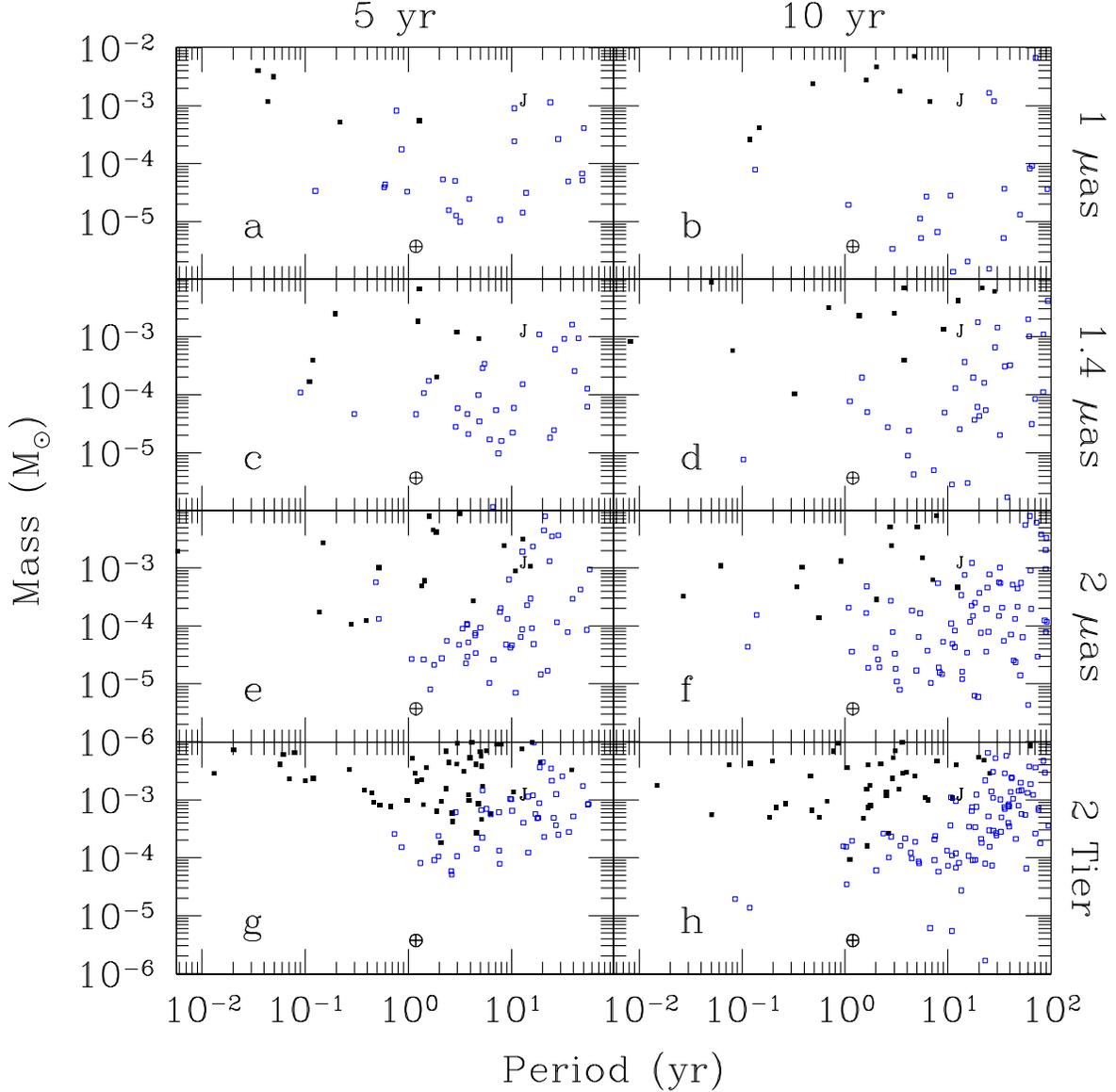}
\caption[Ford.fig7.ps]{
%\small
%
In each panel we show the masses and periods of planets detected
from a single simulation.  The left and right
columns are for five- and ten-year missions, respectively.  The
first, second, and third rows are for 1 $\mu$as, 1.4 $\mu$as, and 2 $\mu$as
single measurement accuracy, respectively.  The bottom row is for
a two-tier target list, including 52 stars at 1
$\mu$as precision and 1092 stars at 4 $\mu$as precision.  In each panel,
open squares and solid squares are for planets detected
by SIM only and by both SIM and
radial velocity searches.  The $\oplus$ and $J$ symbols indicate the masses
and orbital periods of the Earth and Jupiter.
\label{MvsPDetect}}
\normalsize
\end{figure}

\begin{figure}[ht]
\plotone{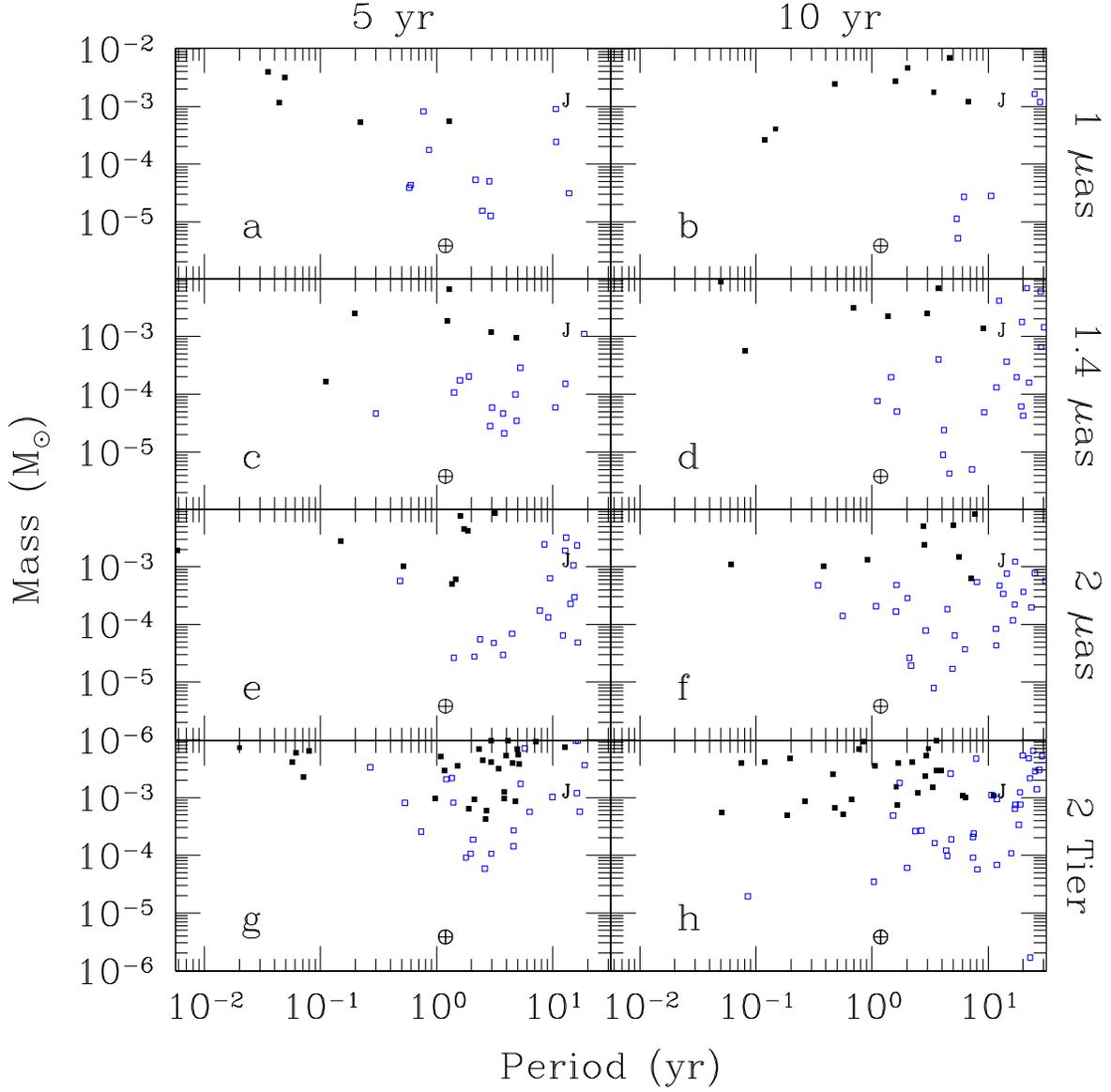}
\caption[Ford.fig8.ps]{
%\small
%
In this figure we show the masses and periods of planets whose masses and orbital
elements are measured with 10\% accuracy.  A single simulation is
shown in each panel.  The left and right columns are for five- and ten-year missions, respectively.
The first, second, and third rows are for 1 $\mu$as, 1.4 $\mu$as, and
2 $\mu$as single measurement accuracy, respectively.  The bottom row
is for a two-tier target list, including 52 stars
at 1 $\mu$as precision and 1092 stars at 4 $\mu$as precision.
In each panel, open squares and solid squares are for planets detected
by SIM only and by both SIM and radial velocity searches.  The
$\oplus$ and $J$ symbols indicate the masses and orbital periods of
the Earth and Jupiter.
\label{MvsPOrbit}}
\normalsize
\end{figure}

In Figs.\ \ref{MvsPDetect} and \ref{MvsPOrbit} we plot realizations of
our Monte Carlo simulations to illustrate the periods and masses that
would be found for various mission parameters.  In Fig.\
\ref{MvsPDetect} we require only detecting the planet.  Of the planets
which are detected in our simulations, roughly half have an orbital
period longer than the mission lifetime and $\sim40\%$ have an orbital
period more than twice as long as the mission lifetime (for both five-
and ten-year missions).  The long
lever arm of large orbits partly compensates for the reduction in
sensitivity due to the larger orbital period.  In Fig.\
\ref{MvsPOrbit} we only plot planets for which masses and orbits are
measured with 10\% accuracy.  These figures illustrate that during a
five-year mission SIM would be able to detect planets in the mass
range $0.01-0.1 M_J$ with orbital periods of several years that are
inaccessible to current radial velocity surveys.  However, SIM is far
more effective when it has the longest possible mission lifetime.  A
ten-year mission is far more likely to detect planets that cannot be
detected by current radial velocity surveys.  Also note that there is
considerable variation between realizations, particularly for lower
mass planets.

% FIGURE 7

% FIGURE 8

\begin{figure}[ht]
\plotone{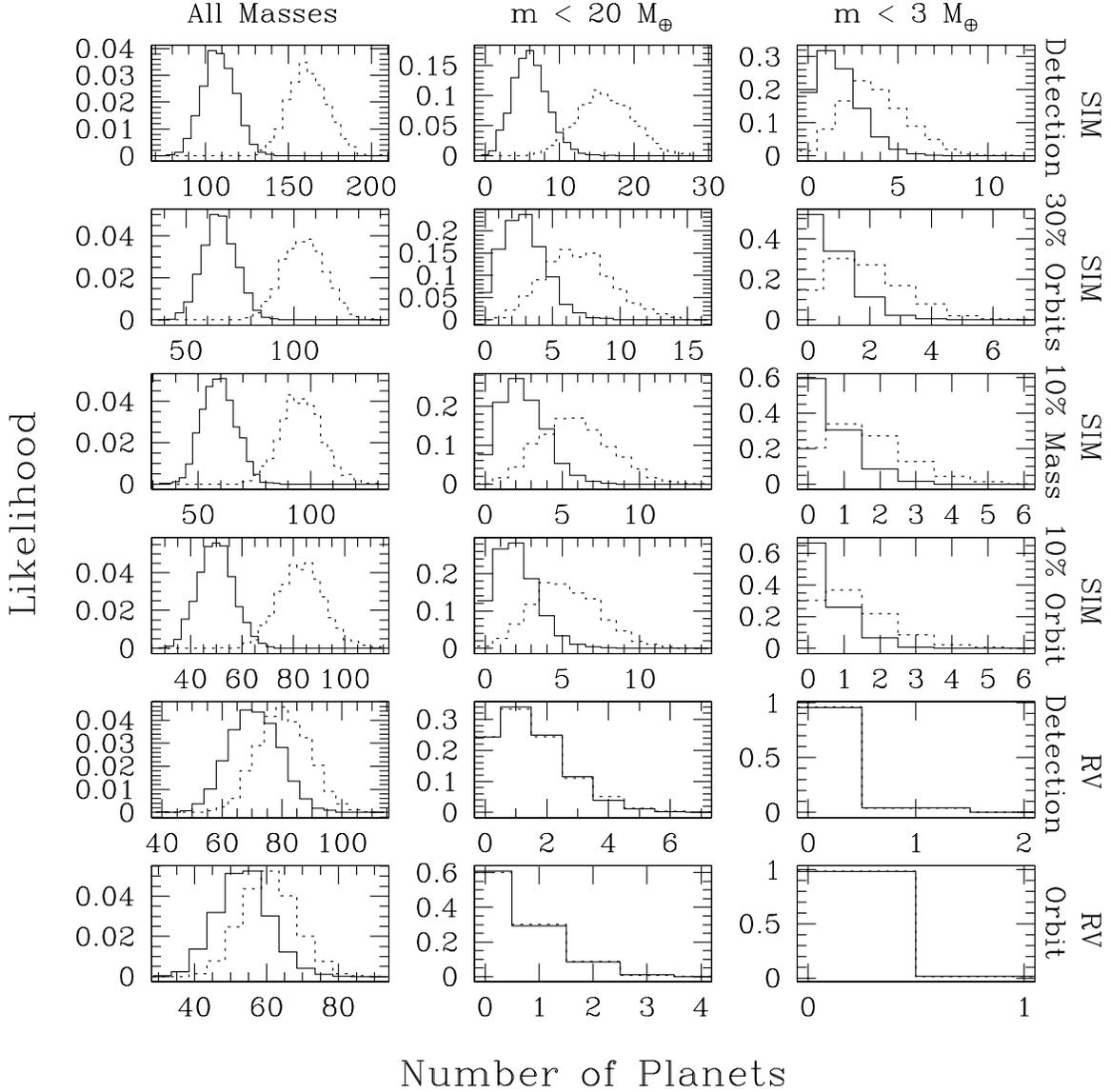}
\caption[Ford.fig9.ps]{
%\small
%
Here we show the likelihood of finding a given number of planets
using a two-tier strategy.
The top four rows are for a SIM survey that targets the 52
nearest stars with a single measurement accuracy of 1 $\mu$as and
1092 conveniently located nearby stars with a single measurement
accuracy of 4 $\mu$as.  
The top row is for detections, the second row is for estimates of
masses and orbital parameters with 30\% accuracy, the third row is for mass
measurements with 10\% accuracy, and the fourth row is for
measurements of the masses and orbital parameters with 10\% accuracy.
The bottom two rows are for a radial velocity survey of the same stars
with 3 \ms single measurement accuracy.  The fifth row is detections
and the bottom row is orbital determinations.
The solid lines are for five-year surveys and the dotted lines are for 10
year surveys.
The left column is for planets of all masses, the center column is for
planets with mass less than $20 M_\oplus$, and the right column is
for planets with mass less than $3 M_\oplus$.
\label{HistogramGridEpics}}
\normalsize
\end{figure}

\begin{figure}[ht]
\plotone{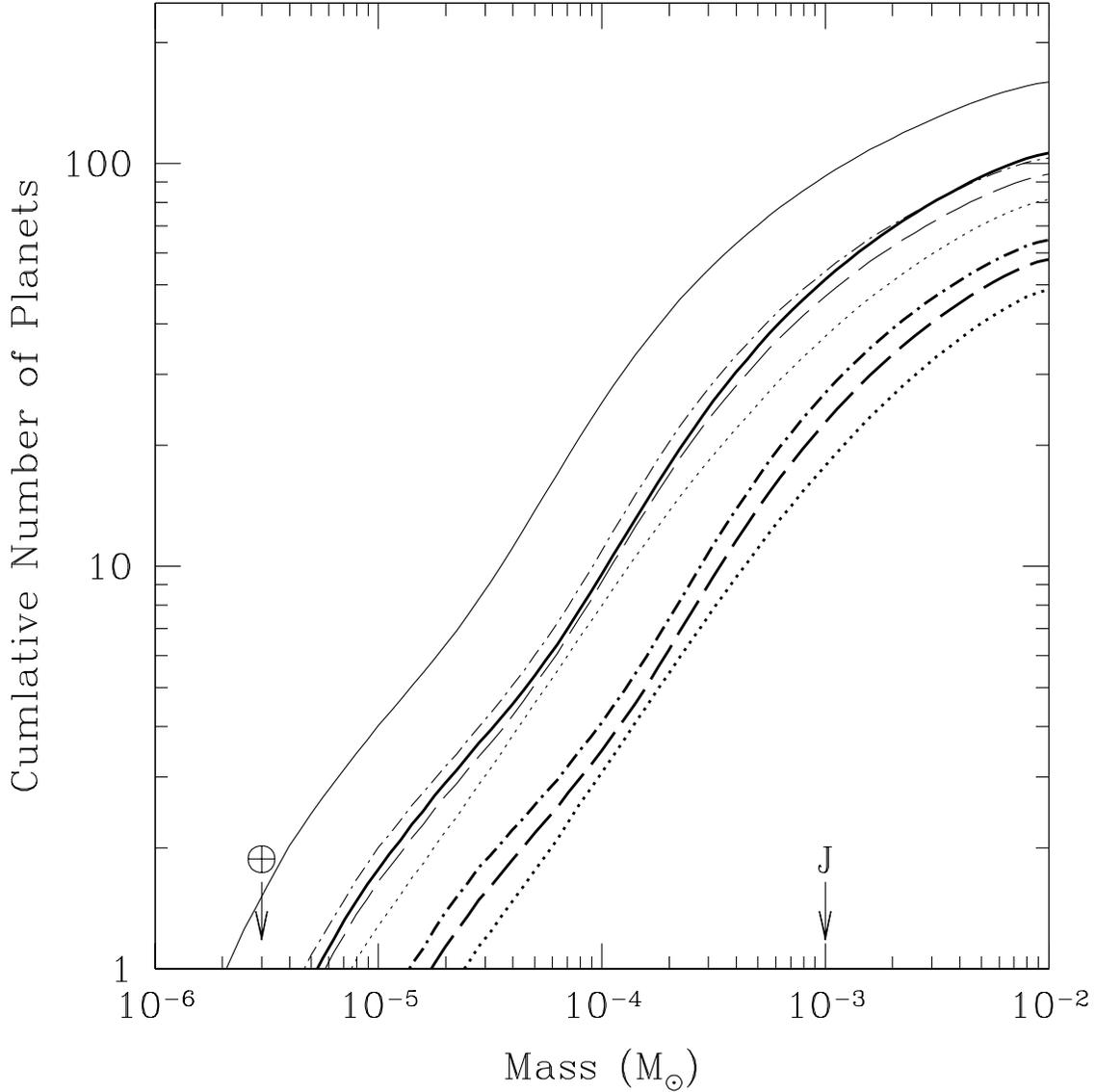}
\caption[Ford.fig10.ps]{
%\small
%
Here we show the results of a Monte Carlo simulation to
estimate how many planets would be found as a function of maximum
mass using a two-tier strategy.  We present the weighted average of thousands of realizations
to yield accurate statistics.
The solid lines are for detections, the dash-dotted lines are for
30\% estimates of orbital parameters, the dashed lines are for 10\%
mass determinations, and the dotted lines are for 10\% orbital
determinations,
The thick lines are for a five-year mission and the thin lines for a ten-year
mission.
These results are for a SIM survey that targets the 52 nearest
stars with a single measurement accuracy of 1 $\mu$as and 1092
conveniently located nearby stars with a single measurement
accuracy of 4 $\mu$as.
The masses of Jupiter and the Earth are indicated by a J and $\oplus$,
respectively.
\label{CumlativeLevelsEpic}}
\normalsize
\end{figure}

In Fig.\ \ref{HistogramGridEpics} and Fig.\ \ref{CumlativeLevelsEpic},
we show the number of planets that a two-tier survey similar to EPIcS
would likely discover and the cumulative distribution of the masses
such a survey would likely detect.  During a five- or ten-year
mission, our two-tier observing strategy would be expected to detect
$\sim108\pm21$ or $162\pm24$ planets and measure masses and orbits
with 10\% accuracy for $\sim50\pm15$ or $84\pm18$ of these, more
planets than would be detected by any of the other SIM observing
strategies that we examined.  During a five or ten year mission, a
two-tier survey would be capable of detecting a planet with a mass
$\sim1.5$ or $0.7 M_\oplus$ and measuring the mass and orbital
parameters with 10\% accuracy for a planet with mass $\sim8 M_\oplus$
or $\sim3 M_\oplus$.  Thus, a two-tier observing strategy similar to
EPIcS could allow SIM to discover and characterize numerous giant
planets, while maintaining a (somewhat reduced) possibility of
detecting terrestrial-mass planets.

% FIGURE 9

% FIGURE 10

\section{Discussion}
\label{SecDiscussion}

In this section we consider several questions about the planet finding
capabilities of SIM in light of our simulations.

\subsection{Will SIM detect Earth-mass planets?}

Of the mission parameters that we consider, a program with $1$ $\mu$as
single measurement accuracy and 120 target stars is most effective for
detecting the lowest mass planets (see Figs.\ 4 and 5).  With these
parameters, it is likely that SIM would detect one or two planets
with $m \le 1 M_\oplus$ in a          ten-year mission, respectively.
While a ten-year mission might measure the mass and orbital parameters
of one $1 M_\oplus$ planet with 30\% accuracy, either mission would be
extremely unlikely to measure either the mass or orbital parameters
with 10\% accuracy.

If the criterion for ``Earth-mass planets'' is relaxed to $m \le 3
M_\oplus$, then $\sim2\pm2$ or $\sim5^{+4}_{-3}$ (90\% confidence
intervals) such planets would be expected to be discovered from a five
or ten-year mission.  
Masses and orbital parameters might be measured with 30\% accuracy for
$1\pm1$ or $2^{+3}_{-2}$ such planets and with 10\% accuracy for $0^{+2}$
or $1^{+3}_{-1}$ such planets.  The
two-tier strategy that we consider would be nearly as effective as a
120 star 1 $\mu$as survey for detecting and measuring orbital
parameters of planets with masses $\le3 M_\oplus$ (see Fig.\ 9).

It is important to note once again that in this mass range, we are
relying on a significant extrapolation of the observed planetary mass
function.  If the actual frequency of Earth-mass planets is
significantly greater than suggested by present radial velocity
observations, then the number of detected Earth-mass planets could be
higher.  If all stars have an Earth-mass planet at 1 AU, a ten-year
survey of 120 stars at $1$ $\mu$as precision could detect $\sim10\pm4$
planets, and measure the masses and orbital parameters with 30\%
accuracy for $\sim3\pm2$ of these. Further, SIM would measure the mass
and orbital parameters to within 10\% for $1^{+2}_{-1}$ such planets.

\subsection{Will SIM detect terrestrial-mass planets?}

Traditional theories of planetary formation (Safronov 1969) predict
that the giant planets form by accretion of gas onto rocky cores of
$5-20 M_\oplus$, while lower mass planets form via a different
mechanism.  Here we evaluate the likelihood for SIM to detect
planets with $m \le 20 M_\oplus$.

If the primary goal is to detect planets with $m< 20 M_\oplus$, then
of the four observing plans we have considered, the one with 2 $\mu$as
single measurement precision is optimal (see Fig.\ 5a) and could be
expected to detect $11\pm6$ or $25\pm9$ planets with masses $\le20
M_\oplus$ for five- and ten-year missions, respectively (see Fig.\ 3
top).  A five- or ten-year mission is expected to make $4^{+4}_{-3}$
or $12\pm6$ mass and orbital parameter determinations with 30\%
accuracy (see Fig.\ 3 middle right) and $3\pm3$ or $8^{+5}_{-4}$ mass
and orbital parameter determinations with 10\% accuracy (see Fig.\ 3
bottom right).

The two-tier strategy would detect $\sim5^{+2}_{-3}$ or $\sim17\pm9$
planets with masses $m\le 20 M_\oplus$ during a five- or ten-year
mission, respectively.  It would measure masses and orbits with
30\% accuracy for $2^{+4}_{-2}$ or $7^{+6}_{-5}$ planets and measure
the masses and orbital parameters with 10\% accuracy for $2^{+3}_{-2}$
or $5^{+5}_{-4}$ such planets depending on the length of the mission.

Again, it is important to remember that these estimates depend on an
extrapolation of the planet mass function observed for larger masses,
although in this case the extrapolation in mass is relatively small,
since radial velocity surveys already detect $\sim30 M_\oplus$ planets.

\subsection{What is the smallest mass planet that SIM is likely to detect?}

Any of the ten-year SIM surveys which we considered is likely to
detect one planet with a mass $\sim1 M_\oplus$, for any of the
target lists we consider.  Similarly, a 1 $\mu$as five-year SIM survey
is likely to detect one planet with a mass $\sim1.5 M_\oplus$ and the
other five-year surveys would likely detect one planet with a mass
$\sim2-3M_\oplus$ (see Figs.\ 4 and 10).  Of course, the mass of the
smallest planet that SIM will actually detect may be significantly
larger or smaller than these predictions due to small number
statistics and uncertainties in the mass function.

\subsection{What is the smallest mass planet that SIM is likely to get orbits for?}

Either a 1, 1.4, or 2 $\mu$as survey is likely to measure with 30\%
accuracy the mass and orbital parameters of one planet with a mass
$\sim5 M_\oplus$ for a five-year survey.  During a ten-year survey,
SIM would likely measure with 30\% accuracy the mass and orbital
parameters of one planet with a mass $\sim 1.5 M_\oplus$ with the
1 $\mu$as survey and $\sim2 M_\oplus$ with either the 1.4 or 2 $\mu$as surveys.
The two-tier strategy would likely measure with 30\% accuracy the mass
and orbital parameters of one planet with a mass $\sim5$ or $1.5
M_\oplus$ for five- or ten-year surveys.

If 10\% accuracy is required, then a 1 $\mu$as survey is likely to
measure the mass and orbital parameters of one planet with a mass
$\sim7$ or $2 M_\oplus$ for five- or ten-year surveys, respectively.
Similarly, a 2 $\mu$as survey is likely to measure with 10\% accuracy
the mass and orbital parameters of one planet with a mass $\sim8$ or
$3 M_\oplus$ for five- or ten-year surveys, respectively (see Fig.\
4).  The two-tier strategy would likely measure the mass and orbital
parameters with 10\% accuracy for a $\sim8 M_\oplus$ planet in five
years and a $\sim3 M_\oplus$ planet in ten years (see Fig.\ 10).

\subsection{How many new planets will SIM detect?}

We can estimate the number of planets of all masses that SIM is likely
to find.  These estimates are much more reliable than the previous
ones, since radial velocity surveys are sensitive to the typical
planets that SIM is expected to find.  Our estimates for a
five-year mission duration are also more reliable than for a ten-year
mission, since radial velocity surveys are just beginning to detect
planets with orbital periods $\sim10$ years.

If its primary goal is to detect and measure masses and orbits for the
greatest number of planets, regardless of mass, then SIM should pursue
a survey of a large number of stars with a relatively low single
measurement accuracy.  For example, using a single measurement
precision of 1 or 2 $\mu$as, SIM would be expected to detect
$\sim27\pm7$ or $69\pm13$ planets in a five-year mission or
$\sim33\pm8$ or $98\pm15$ planets in a ten-year mission (see Fig.\
4).  The two-tier strategy that we consider would likely discover
$\sim108\pm21$ or $162\pm24$ planets for a five- or ten-year mission,
respectively (see Fig.\ 9).

Radial velocity surveys are also capable of discovering giant planets
with orbital periods of several years.  Thus, it is important that we
estimate the number of ``new'' planets which SIM will detect that
would not be discovered by a radial velocity survey of the same stars.
A ten-year radial velocity survey with $3$ \ms precision would be
expected to detect $\sim8\pm4$ and $34\pm9$ planets, for the 1 and 2
$\mu$as target lists.  For the same two target lists, SIM would be
expected to detect $\sim17$ or $44$ new planets in a five-year mission
and $\sim25$ or $68$ new planets in a ten-year mission (see Fig.\ 4).
The two-tier strategy would be expected to detect $\sim57$ or $97$ new
planets for a five- or ten-year mission.  However, conducting such a
large radial velocity survey and including a significant fraction of
M stars would require a large amount of observing time.

\subsection{How many new planets will SIM get orbits for?}

Unfortunately, SIM will not obtain accurate planetary masses or orbits
for many of the planets it detects.  For example, using a single
measurement precision of 1 or 2 $\mu$as, SIM would be expected to
determine masses and orbits with 30\% accuracy for $\sim16\pm6$ or
$44\pm11$ planets for a five-year mission or $\sim22\pm8$ or $66\pm12$
planets for a ten-year mission ($\sim65\%$ of the planets detected
with the same mission parameters).  SIM would be expected to determine
masses and orbits with 10\% accuracy for $\sim13\pm6$ (54\%) or
$36\pm9$ (52\%) planets for a five-year mission or $\sim19\pm7$ (58\%)
or $55\pm11$ (56\%) planets for a ten-year mission, depending on the
single measurement precision.

Ten-year radial velocity surveys of the same stars with 3 \ms accuracy
would determine accurate orbits (but not $m$ and $\sin i$ separately)
for $\sim6\pm4$ or $25\pm8$ planets, for the 1 and 2 $\mu$as target
lists.  Depending on the single measurement accuracy, SIM would be
expected to determine masses and orbits with 30\% accuracy for
$\sim10$ or $26$ new planets (i.e., planets without orbits determined
by the radial velocity survey) for a five year mission and $\sim16$ or
$44$ new planets for a ten-year mission.  The two-tier strategy could
measure masses and orbits with 30\% accuracy for $\sim66\pm16$ and
$106\pm20$ planets for a five or ten year mission, respectively, of
which $\sim31$ and $60$ would not be measured by a radial velocity
survey of the same stars.  Of course, radial velocity measurements
alone cannot determine the mass and inclination separately.

% Similarly, SIM would be expected to determine masses and orbits with 10\% accuracy for $\sim4$ or $8$ new planets for a five-year mission and $\sim8$ or $17$ new planets for a ten-year mission (see Fig.\ 4).

% The two-tier strategy which we consider could measure masses and orbits with 10\% accuracy for $\sim23\pm10$ and $43\pm14$ planets for a five or ten year mission, respectively, of which $\sim4$ and $13$ would not be measured by a radial velocity survey of the same stars.

\subsection{What is the effect of increasing the number of target stars?}

\begin{figure}[ht]
\plotone{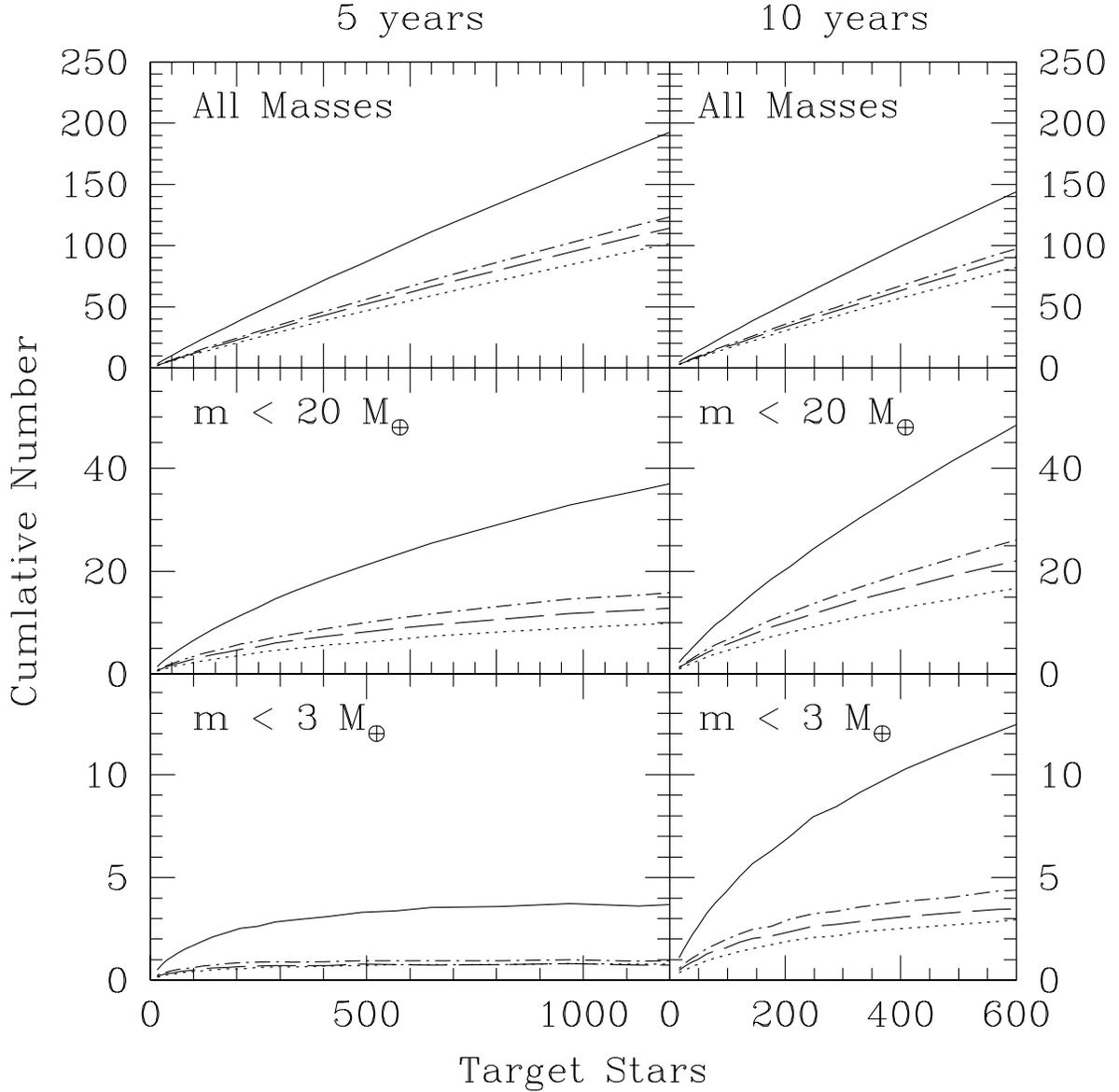}
\caption[Ford.fig11.ps]{
%\small
%
In this figure we plot the number of planets detected as a
function of the number of target stars, assuming an equal number of F,
G, K, and M stars are targeted and $\sigma_d = 1\mu$as.  
The solid lines are for detections, the dash-dotted lines are for
measurements of the mass and orbital parameters with 30\% accuracy,
the dashed lines are for measurements of the mass with 10\% accuracy, and
the dotted lines are for measurements of the orbital parameters with
10\% accuracy.
The top row shows the
number of detections of planets of all masses, the middle row 
shows the number of detections of planets less massive than $20
M_\oplus$, and the bottom row shows the number of detections
of planets with masses less than $3 M_\oplus$.  The left column of
panels is for a five year mission with 24 two dimensional observations,
and the right column of panels is for a ten year mission with 48 two
dimensional observations.  Note that increasing the number of target stars above $\sim300$ in a five-year survey does not substantially increase the number  of $<3 M_\oplus$ planets detected or characterized.
\label{PlanetsVsStars}}
\normalsize
\end{figure}

If additional SIM observing time were to be allocated to planet
surveys, then SIM would be expected to detect a larger number of
planets.  However, the additional target stars would tend to be more
distant than the stars already targeted, so the increase is less than
linear in the number of target stars.  In Fig.\ \ref{PlanetsVsStars}
we plot the number of planets detected (solid lines) as a function of
the number of target stars, assuming an equal number of F, G, K, and M
stars are targeted and $\sigma_d = 1\mu$as.  The dash-dotted lines are
for measurements of the mass and orbital parameters with 30\%
accuracy, the dashed lines are for measurements of the mass with 10\%
accuracy, and the dotted lines are for measurements of the mass and
orbital parameters with 10\% accuracy.  The top row shows the number
of detections of planets of all masses, the middle row shows the
number of detections of planets less massive than $20 M_\oplus$, and
the bottom row shows the number of detections of planets with masses
less than $3 M_\oplus$.  The left column is for a five year mission
with 24 two dimensional observations of each target star, and the
right column is for a ten year mission with 48 two dimensional
observations of each target star.  Clearly, searches for Earth-mass
planets could greatly benefit from a significant increase in the
number of stars targeted.  In principle, a ten year mission could be
subdivided into two five year surveys with separate target lists.
While two separate five year surveys would detect a larger total number of
planets (by $\sim35\%$), a single ten year survey would be expected
to detect slightly more planets with masses less than $20
M_\oplus$ (by $\sim10\%$) and a few times as many planets with masses
less than $3 M_\oplus$.  Therefore, if the SIM mission can be extended
beyond the planned five years, planet searches should continue
observing stars which were part of the original target list.

An alternative strategy for a ten-year mission is to double the number
of target stars and still make only 24 two dimensional observations of
each target star.  This would result in an even greater number of
planets detected and characterized to within 30\% for both planets of
all masses (by $\sim70\%$) and planets with mass less than $20
M_\oplus$ (by $\sim35\%$).  The average number of planets
characterized with masses less than $3 M_\oplus$ would be reduced by
$\sim10\%$ compared to the a ten year survey with twice as many
observations of half as many target stars.

\subsection{How many radial-velocity planets will SIM get orbits for?}

\begin{figure}[ht]
\plotone{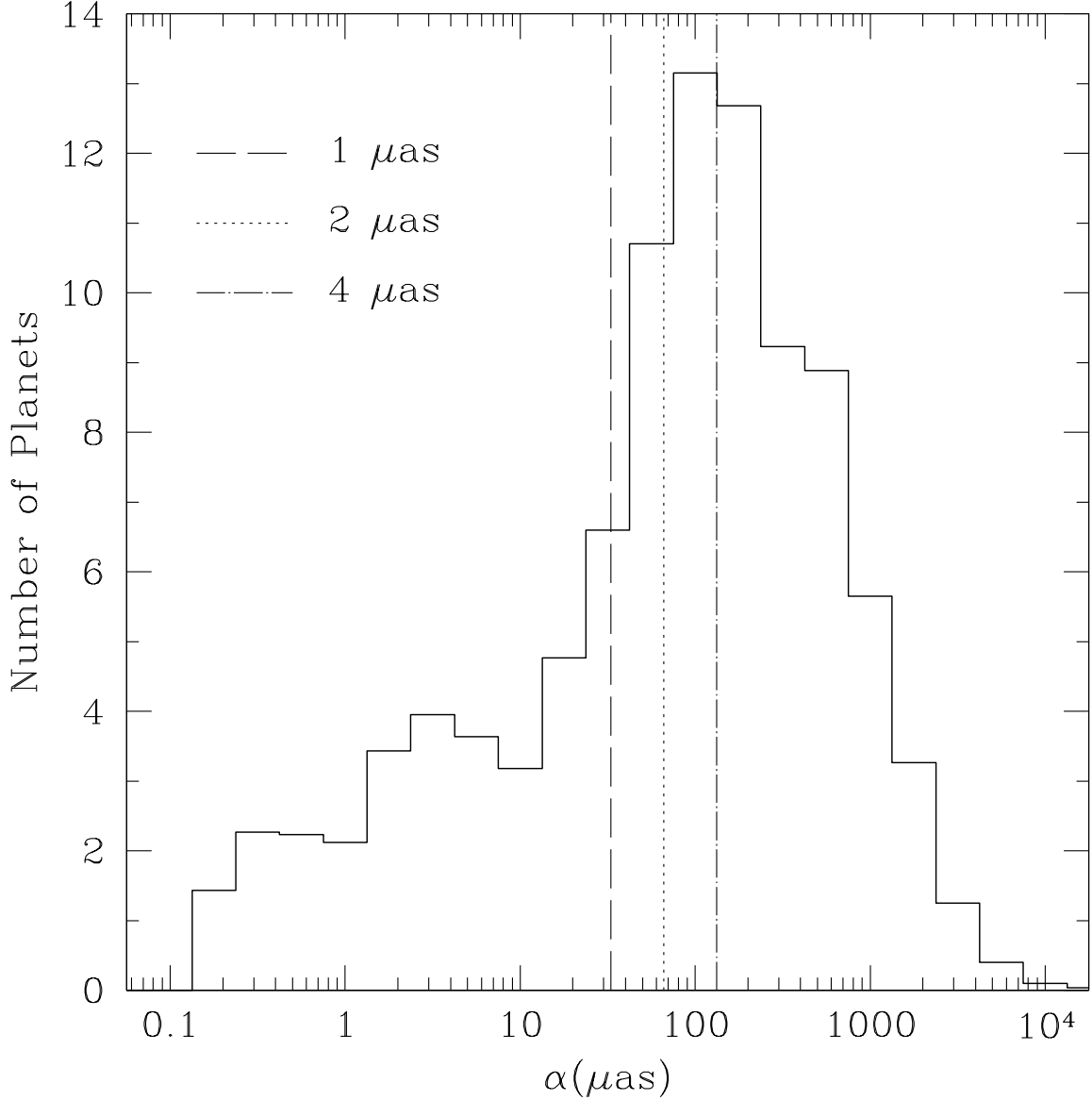}
\caption[Ford.fig12.ps]{
%\small
%
In this figure we show the number of known extrasolar planets with
a given astrometric signal.  The vertical lines show the minimum
astrometric signature necessary for SIM to measure the mass and
orbital elements to within $10\%$ with $95\%$ probability
according to the simulations of Sozzetti \etal (2002).  The
dashed, dotted, and dashed-dotted lines assume single measurement
precisions of 1, 2, and 4 $\mu$as, respectively.  We have assumed
random inclinations of the orbital plane to the sky and averaged
several realizations, resulting in bins that have non-integer
heights.
\label{RvScaledSignal}}
\normalsize
\end{figure}

If SIM were to observe each extrasolar planet
already known from radial velocity surveys, then it would be able to
make 10\% accurate measurements of the mass and orbital elements for $80\pm3$,
$74\pm4$, or $67\pm5$ of the 99 known planets using 24 two dimensional measurements
at $1$, $2$, or 4 $\mu$as precision (see Fig.\ \ref{RvScaledSignal}).
Of the eleven known multiple planet systems, 1 $\mu$as or 2 $\mu$as
astrometry would allow accurate orbital determinations for at least
two planets in seven or five of those systems, assuming that SIM's
ability to measure orbits is not diminished in multiple planet
systems.  In practice, the primary importance of these measurements 
would be to confirm the interpretation of the radial velocity 
measurements and to constrain the inclinations of the orbits.

% FIGURE 11

\section{Conclusions}

All of our results are based on the assumption that the mass function
for terrestrial-mass planets can be obtained by extrapolating the mass
function for giant planets determined by radial velocity surveys.  If
this assumption is correct, then we find that a ten-year 
planet search with SIM might detect a few Earth mass planets.  SIM
could be expected to measure the masses and orbital parameters with
10\% accuracy for planets with masses as small as $\sim2 M_\oplus$.
While SIM could find more low mass planets if they are more common
than current estimates, a null result would not demonstrate that
planets with masses less than $\sim10 M_\oplus$ are less common than
presently expected on the basis of extrapolating the results of radial
velocity surveys.  Including giant planets, SIM is expected to detect
$\sim24-162$ planets; the higher number results from the ten-year
surveys which target a large number of stars at relatively low
precision.  Of these, a significant fraction, $25-65\%$, could also be
discovered by a 3 \ms radial velocity survey.  With the observing
strategies that we have examined SIM will measure the
masses and orbits of $\sim16-106$ planets with 30\% accuracy and the
masses and orbits of $\sim13-84$ planets with 10\% accuracy.  Of
these, $\sim30-70\%$ could be measured by a 3 \ms radial velocity
survey,

\acknowledgments

We thank Stefano Casertano, Jeremy Goodman, Debra Fischer, Geoff
Marcy, Michael Shao, Alessandro Sozzetti, David Spergel, and Edwin
Turner for stimulating discussions.  This research was supported in
part by NASA grant NAG5-10456 and the EPIcS SIM Key Project.

\newpage 

\newpage 
\begin{deluxetable}{lcccc}
\tablecaption{Parameters for detection and measurement probability functions\label{TableFitParam}}
\tablehead{ {Orbital Period}\tablenotemark{a} & {Detection}\tablenotemark{b} & {30\% Mass \& Orbit}\tablenotemark{c} & {10\% Mass}\tablenotemark{c} & {10\% Orbit}\tablenotemark{c} 
}
\startdata
$0$---$0.2$	& 0.77	& 1.77	& 2.81	& 3.11	\\
		& 1.82	& 2.77	& 3.21	& 5.01	\\
		& 2.78	& 1.95	& 8.88	& 9.15	\\
		& 5.59	& 7.09	& 1.95	& 2.14	\\
\\
$0.2$---$0.8$	& 0.86	& 2.03	& 3.10	& 1.24	\\
		& 1.86	& 2.97	& 3.74	& 4.55	\\
		& 2.98	& 1.39	& 7.09	& 3.25	\\
		& 5.34	& 6.58	& 1.43	& 1.24	\\
\\
$0.8$---$1$	& 1.73	& 2.78	& 3.69	& 7.21	\\
		& 1.81	& 3.48	& 4.45	& 8.09	\\
		& 3.62	& 1.31	& 5.69	& 3.11	\\
		& 7.11	& 5.55	& 1.24	& 1.30	\\
\\
$1.$---$1.4$	& 2.25	& 4.15	& 4.34	& 6.70	\\
		& 2.29	& 5.91	& 8.09	& 21.9	\\
		& 5.90	& 6.77	& 6.84	& 4.62	\\
		& 3.25	& 1.40	& 1.01	& 1.41	\\
\\
$1.4$---$2$	& 3.08	& 5.89	& 6.35	& 7.04	\\
		& 3.92	& 11.5	& 16.2	& 47.5	\\
		& 6.12	& 6.85	& 6.31	& 9.57	\\
		& 2.40	& 1.41	& 0.99	& 1.30	\\
\\
$2$---$3$	& 4.37	& 9.08	& 11.5	& 13.4	\\
		& 8.31	& 21.1	& 26.9	& 76.2	\\
		& 7.09	& 6.06	& 4.34	& 5.70	\\
		& 2.12	& 1.59	& 1.22	& 1.27	\\
\\
$3$---$4$	& 11.8	& 11.7	& 27.3	& 42.1	\\
		& 12.7	& 37.3	& 36.0	& 94.6	\\
		& 2.10	& 5.98	& 2.87	& 2.87	\\
		& 2.14	& 1.94	& 1.40	& 1.10	\\
\\
$4$---$8$	& 23.9	& -	& -	& -	\\
		& 38.6	& -	& -	& -	\\
		& 1.75	& -	& -	& -	\\
		& 1.77	& -	& -	& -	\\
\\
$8$---$12$	& 10.8	& -	& -	& -	\\
		& 145.	& -	& -	& -	\\
		& 1.88	& -	& -	& -	\\
		& 1.93	& -	& -	& -	\\
% \\
\enddata
\tablecomments{Each entry lists the parameters $b$, $c$, $\beta$, and $\gamma$ (from top to bottom) for use in equation (5).}
\tablenotetext{a}{Divided by the time span of observations}
\tablenotetext{b}{For a 0.1\% false alarm rate}
\tablenotetext{c}{For 95\% of the planets detected}
\end{deluxetable}

\newpage 
\begin{deluxetable}{lcccccc}
\tablecaption{The number of planets detected or characterized in our simulations\label{TableStatsAll}}
\tablehead{
                     & & 5 Year & & & 10 Year & \\ 
                     & {1 $\mu$as} & {2 $\mu$as} & {2 Tier} & {1 $\mu$as} & {2 $\mu$as } & {2 Tier}
}
\startdata
\qquad {\bf All Masses} & & & & & & \\
Detect               & $24\pm7$ & $69\pm13$ & $108\pm21$ & $33\pm8$ & $98\pm15$ & $162\pm24$ \\ 
Mass \& Orbit (30\%) & $16\pm6$ & $44\pm11$ & $66\pm16$ & $22\pm8$ & $66\pm12$ & $106\pm20$ \\ 
Mass \& Orbit (10\%) & $13\pm6$ & $36\pm9$ & $50\pm15$ & $19\pm7$ & $55\pm11$ & $84\pm18$ \\ 
\tableline 
\qquad $\mathbf{m \le 20 M_\oplus}$ & & & & & & \\
Detect               & $7\pm5$ & $11\pm6$ & $5^{+2}_{-3}$ & $13\pm6$ & $25\pm9$ & $17\pm9$ \\ 
Mass \& Orbit (30\%) & $4\pm3$ & $4^{+4}_{-3}$ & $2^{+4}_{-2}$ & $8\pm4$ & $12\pm6$ & $7^{+6}_{-5}$ \\ 
Mass \& Orbit (10\%) & $2^{+3}_{-2}$ & $3\pm3$ & $2^{+3}_{-2}$ & $5^{+4}_{-3}$ & $8^{+5}_{-4}$ & $5^{+5}_{-4}$ \\ 
\tableline 
\qquad $\mathbf{m \le 3 M_\oplus}$ & & & & & & \\
Detect               & $2\pm2$ & $1^{+2}_{-1}$ & $1^{+3}_{-1}$ & $5^{+4}_{-3}$ & $4^{+4}_{-3}$ & $3^{+5}_{-2}$ \\ 
Mass \& Orbit (30\%) & $1\pm1$ & $0^{+1}$ & $0^{+2}$ & $2^{+3}_{-2}$ & $1^{+3}_{-1}$ & $2\pm2$ \\ 
Mass \& Orbit (10\%) & $0^{+2}$ & $0^{+1}$ & $0^{+2}$ & $1^{+3}_{-1}$ & $1^{+2}_{-1}$ & $1^{+2}_{-1}$ \\ 
\enddata
\tablecomments{Error bars correspond to 90\% confidence intervals.}
\end{deluxetable}

\end{document}